Thermodynamic modeling using Extended UNIQUAC and COSMO-RS-ES models: Case study of the cesium nitrate - water system over a large range of temperatures.

Mouad Arrad, Department of Process Engineering, Equipe Énergie et Procédés Propres, National School of Mines of Rabat, BP 753 Agdal, Rabat, Morocco, ORCID: 0000-0001-9752-0501

Kaj Thomsen, Center for Energy Resources Engineering, (CERE), Department of Chemical and Biochemical Engineering, Technical University of Denmark, Kongens Lyngby 2800, Denmark, ORCID: 0000-0003-1373-1630

Simon Müller, Institute of Thermal Separation Processes, Hamburg University of Technology, Hamburg, Germany, ORCID: 0000-0003-1684-6994

Irina Smirnova, Institute of Thermal Separation Processes, Hamburg University of Technology, Hamburg, Germany, ORCID: 0000-0003-4503-4039

Abstract

A comparison of two thermodynamic models is presented using the water-cesium nitrate system as case study. Both models were able to model the thermodynamic properties such as the osmotic coefficient, vapor pressure, mean activity coefficient and solubility with good accuracy. We show that it is possible to reproduce the temperature dependency of the properties using a simple set of parameters in the case of Extended UNIQUAC. Furthermore, COSMO-RS-ES is a completely predictive model adjusted to data at 298.15 K, which is applied for the first time to other temperatures.

**Introduction**

Cesium nitrate is a highly water-soluble crystalline cesium source for uses compatible with nitrates and lower (acidic) pH. Nitrate compounds are generally soluble in water. Nitrate materials are also oxidizing agents. When mixed with hydrocarbons, nitrate compounds can form a flammable mixture. Nitrates are excellent precursors for production of ultra-high purity compounds. Certain catalysts and nanoscale products (nanoparticles and nano-powders) are based on cesium nitrate salts [1-2]. It is used in pyrotechnic compositions, as a colorant and an oxidizer, e.g. in decoys and illumination flares. Cesium nitrate prisms are used in infrared spectroscopy, in x-ray phosphor and in scintillation counters. [3-4] It is also used in making optical glasses and lenses.

As a rare essential metal, cesium and its compounds are widely used in aerospace, electronic devices, biomedicine, and other modern technical fields [4]. The reserves of nonradioactive cesium resources in Salt Lake brine, geothermal water, and underground brines are abundant [5-7].

The study of the thermodynamic properties for this system is very important regarding the wide area of use and the valuable information that modeling can bring to the scientific community.

Previously, the ternary $CsNO_3$-$NH_4NO_3$-$H_2O$ system was modeled [8] using the Pitzer model extended with Harvie-Weare equations. The focus was to reproduce the $CsNO_3$-$NH_4NO_3$-$H_2O$ phase diagram at 298.15 K and 348.15 K measured by Li et al.[9] This modeling required a total of 20 parameters [8].

In this work a comparison of two different categories of thermodynamic models was carried out. The Extended UNIQUAC and the COSMO RS-ES models are widely different but are here applied to the same problem. The main outcome of this work is to see how easily it is to widen the applicability of Extended UNIQUAC to the binary system $CsNO_3$ and test how well the predictive model COSMO-RS-ES can be applied to systems at temperatures different from 298.15 K.

**Analysis of available experimental data:**

The experimental data of the water-cesium nitrate system are relatively scarce. The work of Robinson [10] is the first set of osmotic coefficient and activity coefficient to be reported at concentrations up to 1.5 molal at 298.15 K. Fanghänel et al.[11] measured the osmotic coefficient at 373.25 K from 2.05 to 7.21 molal. Frolov et al., [12] reported water activities at 298.15 K. Ryabov et al.[13], measured the water activity at 298.15 K of one $CsNO_3$ solution. Kirgintsev et al. [14] reported the water activity of two $CsNO_3$ solutions at 298.15 K. Harned and Owen [15] measured the mean activity coefficient at very low concentration at 298.15 K using the emf method.

Richards and Rowe [16], reported a few heat of dilution data and heat capacity data at 293.15 K. Jauch [17] reported heat capacity data at 291.15 K. The amount of thermal property data available was considered insufficient for parameter estimation. These data points were therefore not included in this modeling project.

Freezing point data were reported by Washburn and Macinnes [18] and Roth-Greifswald [19]. Apelblat and Korin [20] measured vapor pressure data at several temperatures from 280.95 K to 323.55 K.

Solubility values were determined by Berkeley [21], Yakimov et al. [22], Arkhipov et al. [23], Cherkasov et al. [24-25], and recently by Li et al, [8].

The available experimental data are relatively limited but a good agreement between data from different sources and different types of data creates a consistent view of the thermodynamic behavior.

The experimental data used for modeling in this project are listed in Table 1.

Table 1: Available data from the literature used for determining model parameters for the cesium nitrate-water system

| Data type | Temperature (K) | Molality | Data points | Reference |
|---|---|---|---|---|
| Osmotic coefficient | 298.15 | 0.01 - 1.41 | 60 | Robinson [10] |
| Activity coefficient | 298.15 | 0.0005 - 0.01 | 5 | Harned and Owen [15] |
|  | 298.15 | 0.435 - 1.570 | 8 | Frolov et al.[12] |
|  | 298.15 | 3.1 | 1 | Ryabov et al. [12] |
| Vapor pressure | 298.15 | 1.087 - 1.241 | 2 | Kirgintsev and Luk'yanov [14] |
|  | 373.45 | 1.115 - 7.212 | 13 | Fanghänel et al. [11] |
|  | 280.95 - 323.55 | 0.709 - 3.302 | 28 | Apelblat and Korin [20] |
| Freezing – SLE | 271.89 - 273.08 | 0.018 - 0.437 | 11 | Washburn and Macinnes [18] |
| Salt solubility | 278.15 - 393.15 | 0.608 - 13.942 | 15 | Cherkasov et al.[24-25] |

Modeling approach:

Extended UNIQUAC model

The Extended UNIQUAC model is a thermodynamic model for electrolyte solutions. It consists of a UNIQUAC term [26] and an Extended Debye-Hückel term. The model was developed over several years and is used in its current form here as it was presented by Thomsen et al. [27]

The Extended UNIQUAC model is mole fraction based and uses UNIQUAC interaction parameters with linear temperature dependence [27]. It therefore offers the possibility to model electrolyte solutions in the entire concentration range and over wide temperature ranges. The Extended UNIQUAC model is a good option for modeling solid-liquid equilibrium, vapor-liquid equilibrium, and various properties of aqueous salt solutions.

A significant advantage of the Extended UNIQUAC model compared to some other models is that temperature dependence is built into the model. This enables the model to also describe thermodynamic properties that are temperature derivatives of the excess Gibbs energy, such as heat of mixing and heat capacity.

For modeling the binary $CsNO_3$ – $H_2O$ system, the nitrate parameters reported by Thomsen et al.[27] were used unchanged. UNIQUAC volume and surface area parameters r, and q for the cesium ion were determined in this work. The volume and surface area parameters for ions in aqueous solutions are considered adjustable parameters in this model. Ideally, the volume and surface area parameters are determined from data for several binary and ternary aqueous solutions with several different ions. In this case, some solid-liquid equilibrium data [28-32] for the $Cs_2CO_3$-$H_2O$ system were included together with data for the $CsNO_3$-$H_2O$ system to determine these parameters. The r parameter for $Cs^+$ was determined to be 5.75 and the q parameter for $Cs^+$ was determined to be 1.51.

In addition to the volume and surface area parameters, the interaction parameter for the interaction of $Cs^+$ with $NO_3^-$ was determined. This temperature dependent interaction parameter consists of a basis value and a temperature gradient. The basis value was determined to be 564.652 and the temperature gradient was determined to be 2.4287. The interaction parameter can therefore be written: $u_{Cs-NO3} = 564.652 + 2.4287 * (T - 298.15)$.

The model parameters were determined by minimization of the weighted difference between calculated and experimental data using a modified Marquardt subroutine for non-linear least squares from the Harwell subroutine library [33].

COSMO-RS-ES

COSMO-RS-ES [34-35] is a predictive model which, like Extended UNIQUAC, consists of a short-range and a long-range term. It combines a tailored version of open COSMO-RS [36] and an especially improved long-range term ME-PDH [37]. The model has seen several improvements over the years [38-39] leading to a model of broad predictive applicability for electrolyte solutions in mixed solvents up to very high salt concentrations. Even for completely non-aqueous systems and for low permittivity systems, the model is able to capture trends semi quantitatively without the need of any type of binary interaction parameter. The improved long-range model has been shown to be able to describe systems including ILs better than conventional terms [40]. So far, the model has been only applied to electrolyte systems at 298.15 K which is why the application in this work is considered purely predictive. Ongoing work will include a temperature dependency to improve the model further.

Results and discussion:

The modeling results show a good agreement between the calculated osmotic coefficient from the Extended UNIQUAC model and the COSMO-RS-ES model at

298.15 K. Water activity data shows a deviation for the COSMO-RS-ES starting from 1 molal.

Water activity at around 373.15 K is correctly calculated by Extended UNIQUAC, while COSMO-RS-ES predicts the correct trend with some deviations as the salt concentration increases.

Figure 1, Figure 2, and Figure 3 show calculated and experimental values of osmotic coefficient and water activity.

The prediction of the mean activity coefficient follow the same trend in both models at 298.15 K. Figure 4 shows the mean activity coefficient predicted by the models.

Both models are in excellent agreement with vapor pressure data for saturated solutions measured by Apelblat and Korin [20]. This is shown in figure 5. The good agreement of both models is expected, as the vapor pressure is not that sensitive towards the salt concentration.

Concerning the solubility, Extended UNIQUAC was able to reproduce the solubility phase diagram with high precision including the ice curve and the branch for anhydrous cesium nitrate using the solubility product calculated from standard state properties from the NBS tables [41-43]. In the Extended UNIQUAC model, the Gibbs-Helmholtz equation is used for determining the solubility product at temperatures different from 298.15 K [44]. Figure 6 shows the calculated diagram along with the experimental data from the literature. For COSMO-RS-ES the solubility calculation did not converge using the solubility product calculated from the NBS tables. This is also expected as the osmotic coefficient predicted by the model at 298.15 K deviates from the experimental value (Fig. 1). For this reason, the solubility product for COSMO-RS-ES was calculated from the activity at the experimental concentration at 298.15 K and then used to predict the solubility at other temperatures. Even though the model has no temperature dependency for electrolyte systems and was only adjusted to experimental data at 298.15 K, it reproduces relatively well the temperature dependence of the solubility in a range of 283.15 to 318.15 K.

………………………………………………………………….

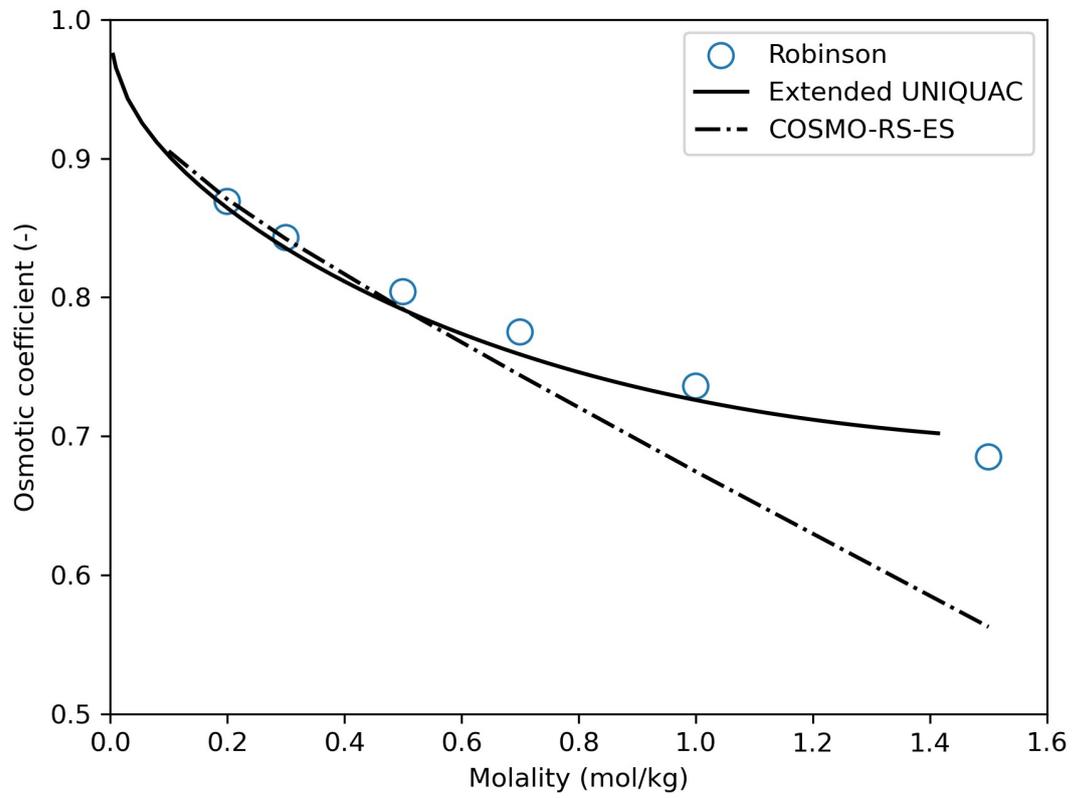

Figure1 : Osmotic coefficient from Robinson [10] and values calculated using both models at 298.15 K

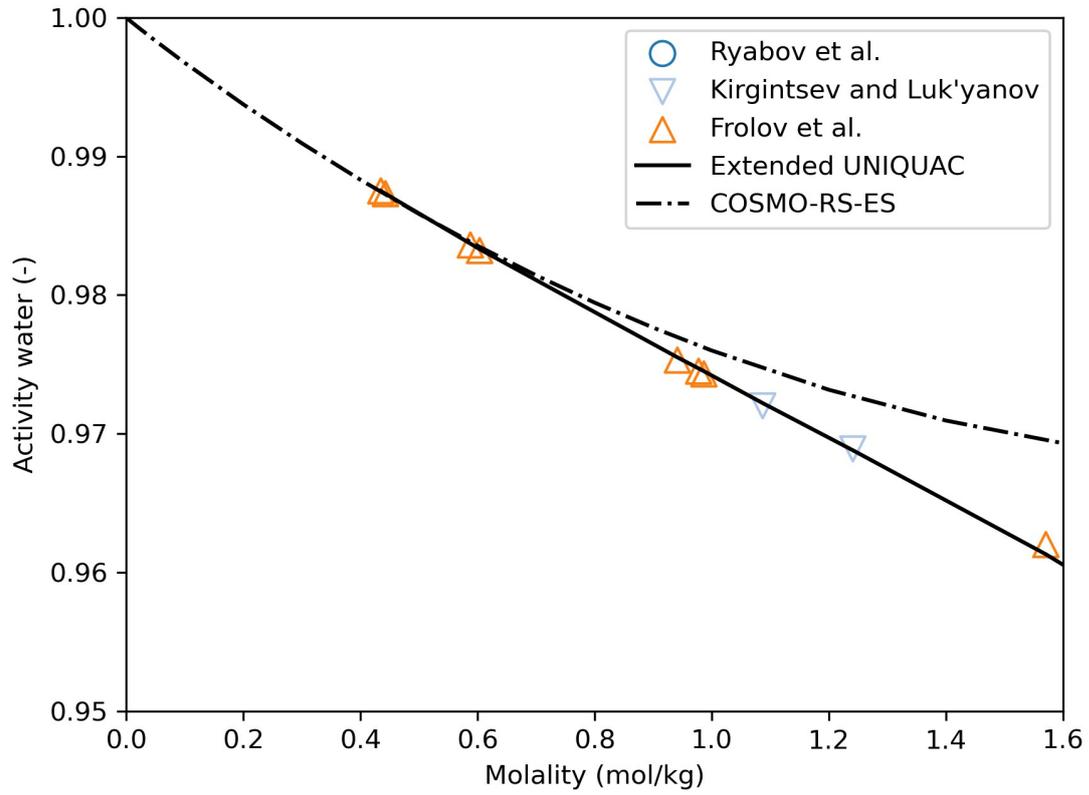

Figure 2 : Experimental water activity plotted with values calculated using COSMO-RS-ES and Extended UNIQUAC models at 298.15 K

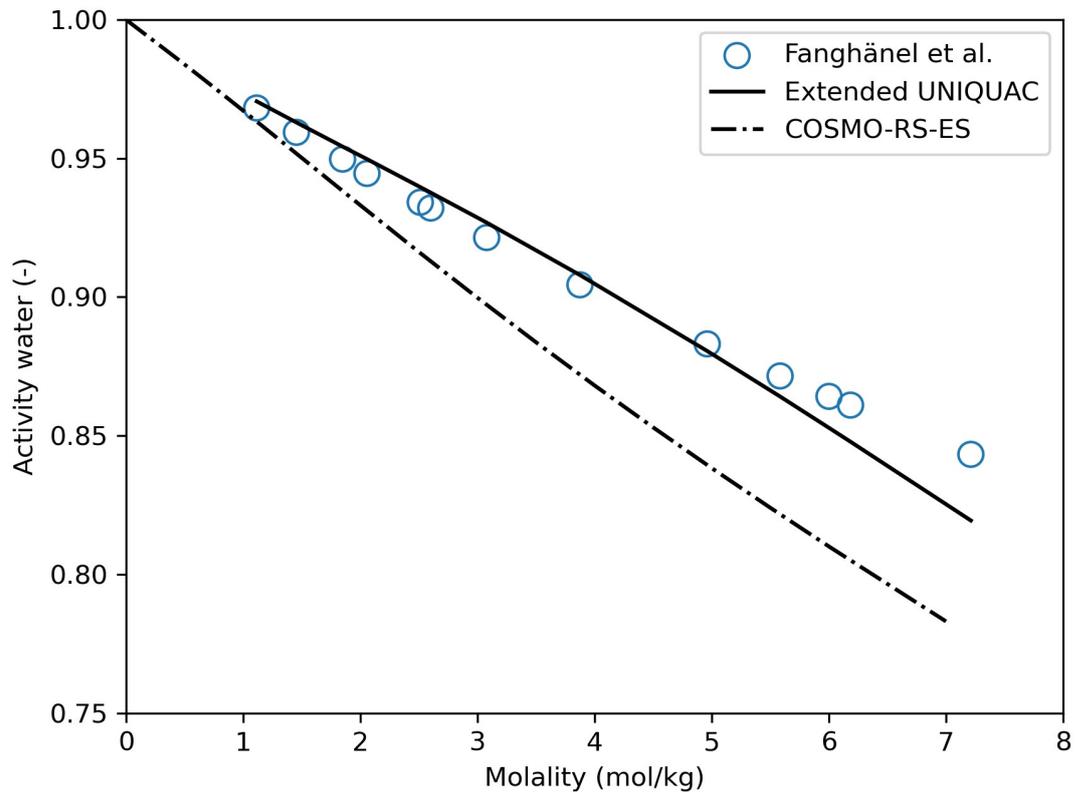

Figure 3 : Experimental water activity calculated using COSMO-RS-ES and Extended UNIQUAC models at 373.45 K

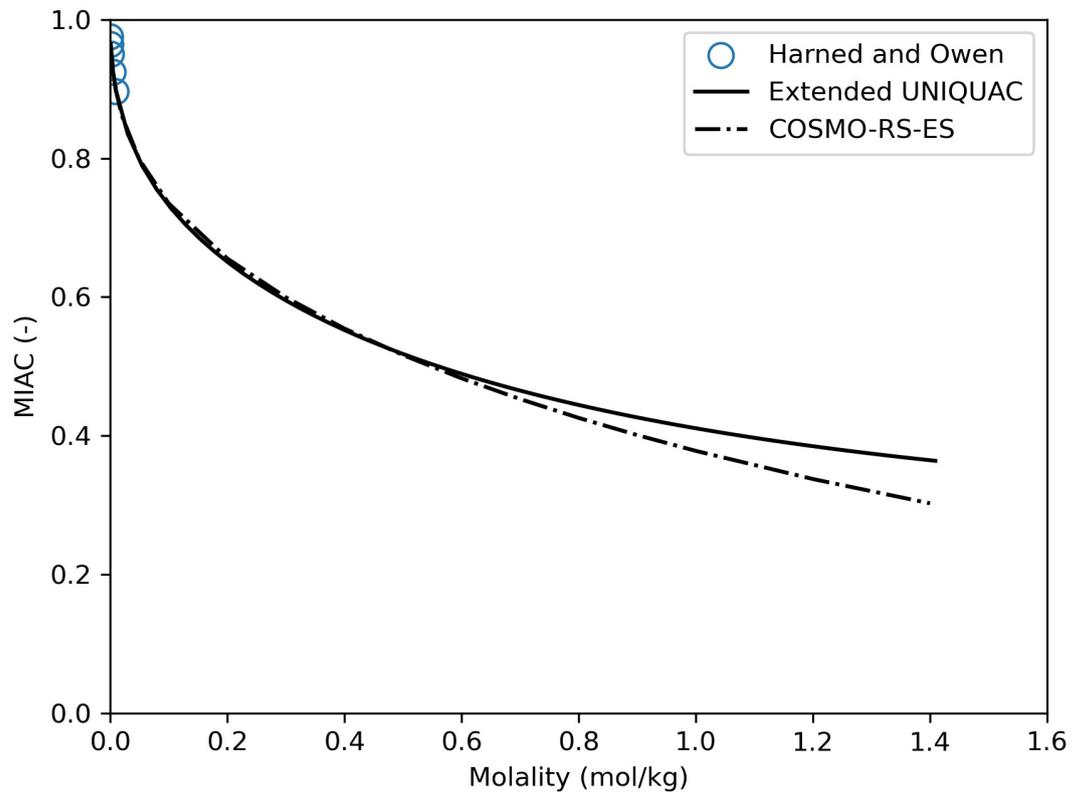

Figure 4 : Mean activity coefficient calculated using both models and low concentration data from Harned and Owen [15]

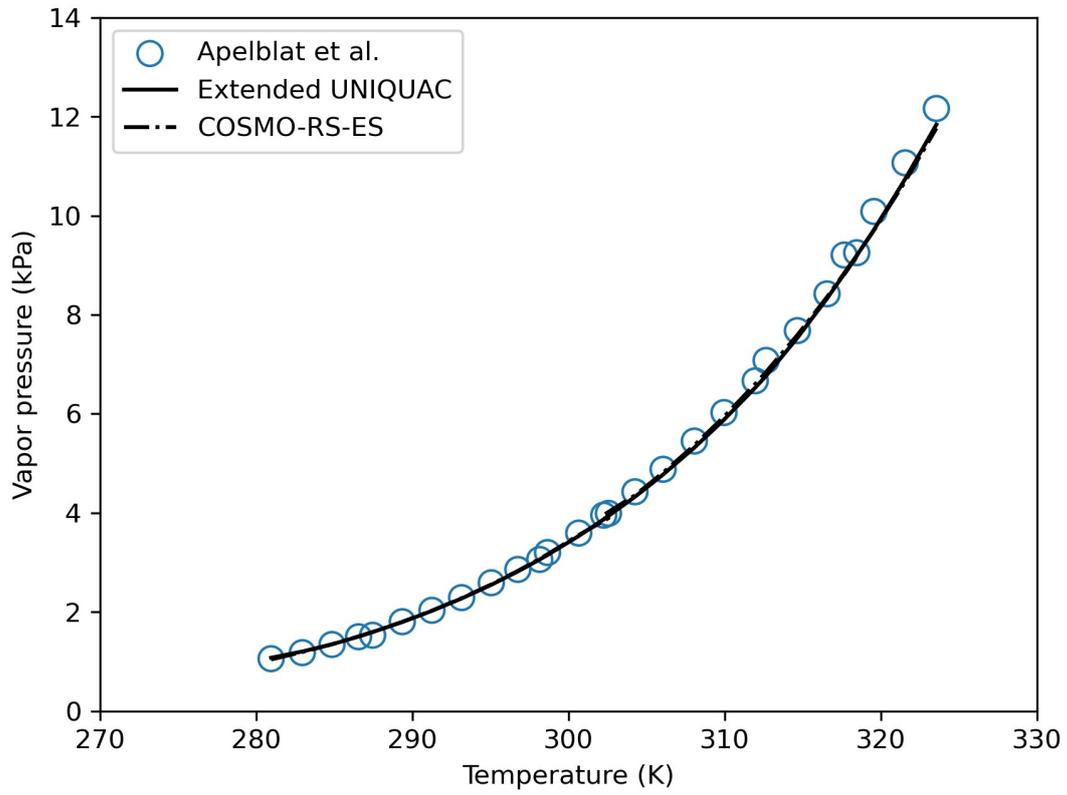

Figure 5 : Experimental vapor pressure data at saturation and calculated vapor pressures at saturation using both models in the range 280-323.15 K

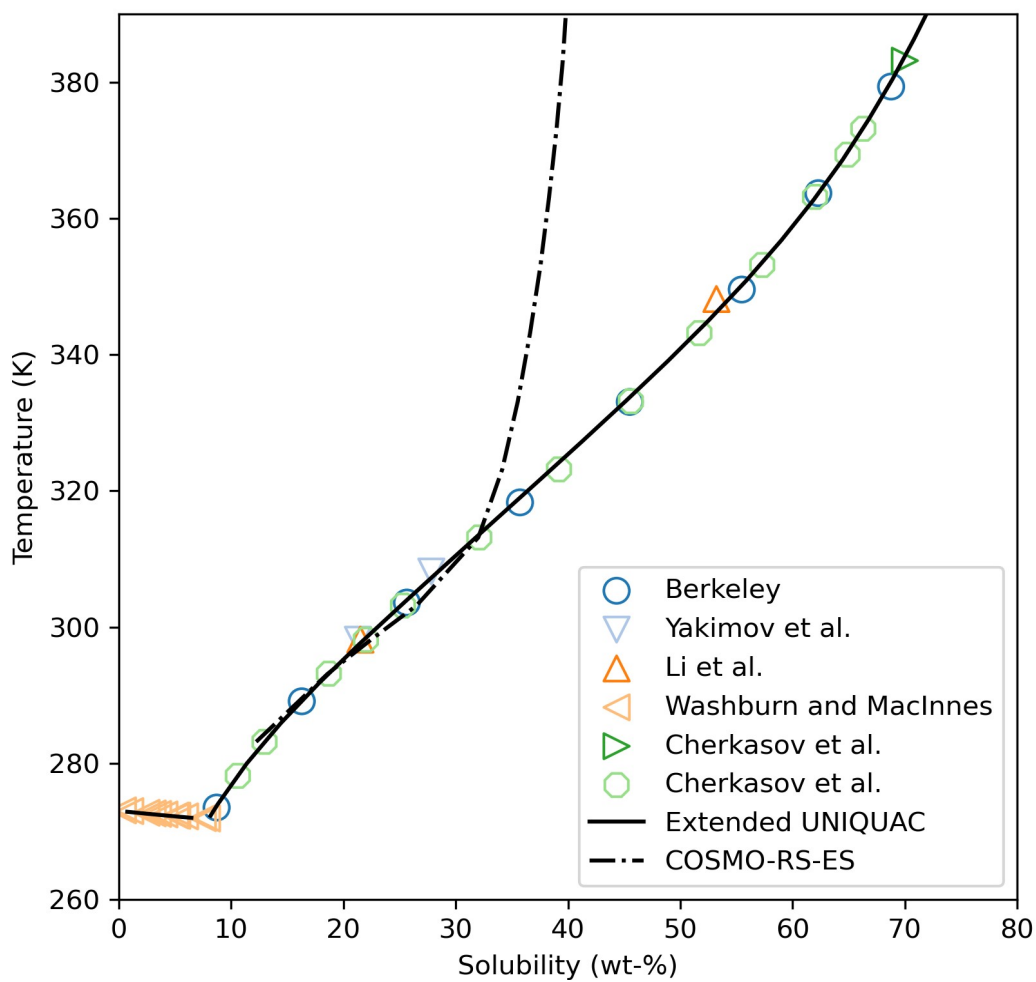

Figure 6 : Solubility calculated with both models and available experimental solubility data plotted on the graph.

Conclusions

The goal of this work was to widen the applicability of Extended UNIQUAC to the system of water + $CsNO_3$ and apply COSMO-RS-ES to temperature dependent data for the first time. The models were able to predict/reproduce data for osmotic coefficients, vapor pressures, and activity coefficients quite well. The Extended UNIQUAC model was able to reproduce the solubility of the water-cesium nitrate system very accurately. COSMO-RS-ES shows its predictive capabilities, being a model without any binary interaction parameters calculating the thermodynamics purely based on ab-initio information. In current works on COSMO-RS-ES, improvements for the temperature dependence in electrolyte systems are being explored.